\documentclass{aa3}
\usepackage{graphicx}

\begin{document}

\title{Star-forming protoclusters associated with methanol masers}

   \author{V. Minier
              \inst{1,2}
          \and M.G. Burton
          \inst{2}
          \and T. Hill
          \inst{2}
          \and M.R. Pestalozzi
          \inst{3}
          \and C.R. Purcell
          \inst{2}
          \and G. Garay
          \inst{4}
          \and A. Walsh
          \inst{5}
          \and \\ S. Longmore
          \inst{2}}

   \offprints{Vincent Minier }

\institute{Service d'Astrophysique, DAPNIA/DSM/CEA CE de Saclay, 91191 Gif-sur-Yvette, France \\
              \email{Vincent.Minier@cea.fr}
  \and School of Physics, University of New South Wales, Sydney 2052, NSW, Australia
  \and Onsala Space Observatory, S-439 92 Onsala, Sweden
  \and Departamento de Astronom\'{\i}a, Universidad de Chile, Casilla 36-D, Santiago, Chile
  \and Center for Astrophysics, 60 Garden Street, Cambridge, MA, 02138, USA \\
}

   \date{Received / Accepted}

   \abstract{We present a multiwavelength study of five methanol maser sites which are not directly associated with a strong ($>100$ mJy) radio
   continuum source: G\,31.28+0.06, G\,59.78+0.06, G\,173.49+2.42 (S231, S233IR), G\,188.95+0.89 (S252, AFGL5180) and G\,192.60-0.05 (S255IR).
   These radio-quiet methanol maser sites are often interpreted as precursors of ultra-compact \ion{H}{ii} regions or massive
   protostar sites. In this work, the environment of methanol masers is probed from mid-IR to millimetre wavelengths at angular
   resolutions of $8''-34''$. Spectral energy distribution (SED) diagrams for each site are presented, together with mass and 
   luminosity estimates.
   Each radio-quiet maser site is always associated with a massive ($>50$ M$_{\odot}$), deeply embedded ($A_v>40$~mag) and very luminous
   ($>10^4$~L$_{\odot}$) molecular clump, with $L_{total}{\propto}M_{gas}^{0.75}$. These physical properties characterise massive 
   star-forming clumps in earlier evolutionary phases
   than \ion{H}{ii} regions. In addition, colder gas clumps seen only at mm-wavelengths are also found near the methanol maser sites. These colder 
   clumps may represent an even earlier phase of massive star formation. These results suggest an evolutionary sequence for massive star
   formation from a cold clump, seen only at mm wavelengths, evolving to a hot molecular core with a two-component SED with peaks
   at far-IR and mid-IR wavelengths, to an (ultra-compact) \ion{H}{ii} region. Alternatively, the cold clumps might be
   clusters of low-mass YSOs, in formation near the massive star-forming clusters. Finally, the values of the dust grain emissivity index ($\beta$)
   range between 1.6 and 1.9.
     }

\titlerunning{Star-forming protoclusters associated with methanol masers}

\authorrunning{V. Minier et al.}

\maketitle

\section{Introduction}

Class II methanol masers were originally detected toward regions of massive star formation (MSF) because many searches at 6.7 and 12.2 GHz
had focused on observing regions exhibiting typical MSF signatures, such as strong radio continuum sources (i.e. \ion{H}{ii} regions), IRAS sources
with IR colours of ultra-compact \ion{H}{ii} (UC \ion{H}{ii}) regions\footnote{In this paper, a UC \ion{H}{ii} region is defined as a region 
ionised by high-mass stars, with a diameter $\sim0.1$~pc and an electron density ${\ge}10^4$~cm$^{-3}$ producing a 22-GHz flux of 
$\sim100$~mJy at a distance of 5 kpc (Kurtz et al. \cite{kurtz00}).} and OH masers (e.g. Batrla et al. \cite{batrla87}; Schutte et al. 
\cite{schutte93}; Caswell et al. \cite{caswell95}). At low angular resolution ($\sim1'-10'$), many 6.7 and 12.2-GHz methanol masers 
appeared to be associated with \ion{H}{ii} regions.

In contrast, high angular resolution ($\sim0.01''-1''$) observations have demonstrated that methanol masers are in
fact generally not co-spatial with strong radio sources (e.g. UC \ion{H}{ii} regions), but frequently tend to be {\it isolated} from them. Out of
$\sim250$ methanol maser sites studied, 75\% are not directly associated (within 0.1 pc) with UC \ion{H}{ii} regions
or more extended radio sources (Caswell \cite{caswell96}; Walsh et al. \cite{walsh98}; Phillips et al. \cite{phillips98};
Minier et al. \cite{minier01}). Very weak ($<4$ mJy) radio continuum emission has, however, been detected toward isolated methanol maser 
sites (e.g. van der Walt et al. \cite{walt03}). An extreme case is IRAS\,20126+4104 where the methanol maser in the hot molecular core is 
associated with a 0.1-mJy radio source at 8.6~GHz (Hofner et al. \cite{hofner99}; Minier et al. \cite{minier01}).
This might be the first sign that some isolated methanol masers are associated with precursors of UC \ion{H}{ii} regions, i.e. protostellar objects
on the way to forming a star or a multiple star system. Methanol maser sites are therefore generally {\it radio-quiet} in terms of continuum
emission ($<1$ mJy at 5 kpc, i.e. a typical observational sensitivity limit).

Complementary spectral line observations confirm this view. Preliminary results from a molecular line survey with the Mopra millimetre telescope
suggest that isolated methanol masers are possibly associated with hot molecular cores (HMCs) (Purcell et al. \cite{purcell04}), which are presumed
to be the sites where massive protostars evolve to form an ionising ZAMS star (Kurtz et al. 2000; Fontani et al. 2003). The intense and
complex chemistry, the temperature (some 100\,K) and the density (${\ge}10^7$~cm$^{-3}$) of these HMCs would provide suitable conditions
for methanol masers to arise (see Minier et al. 2003 and references therein).

Finally, recent ATCA observations by Minier et al. (\cite{minier03}) might indicate an exclusive association between methanol masers
and massive star-forming regions following a nil detection rate of masers toward regions of low-mass star formation.
All these elements tend to demonstrate that methanol masers may trace an early stage of MSF. Whether methanol masers
trace embedded massive protostars or suitable conditions nearby a site of MSF remains an open question.

This paper is devoted to a multiwavelength
study of five radio-quiet methanol maser sites and their environments, which are particularly good candidates for hosting a massive protostar
(Minier et al. \cite{minier00}, \cite{minier01}): G\,31.28+0.06, G\,59.78+0.06,
G\,173.49+2.42 (S231, S233IR), G\,188.95+0.89 (S252, AFGL\,5180) and G\,192.60-0.05 (S255IR).
Comparisons of images taken from optical to radio wavelengths
are presented, accompanied by estimates of the mass, luminosity and temperature of the MSF clumps in each region. We will
demonstrate that the five radio-quiet maser sites coincide with massive and luminous star-forming clumps. Finally, it will be
shown that methanol masers trace very young clusters of star formation in the earliest stages of their evolution.

Sect. 2 describes the observations and data archives used for this work. Derived physical quantities and diagrams of spectral energy
distributions are presented in Sect. 3, followed by a discussion in Sect. 4 and conclusions in Sect. 5.

\section[]{Observations, data archives and analysis}

\subsection{Optical image archives}

An optical image of an $8'\times8'$ region was generated around each source using the Digitized Sky Survey (DSS) database
(http://archive.eso.org/dss/dss). The Digitized Sky Survey comprises a set of all-sky photographic surveys conducted with the Palomar and
UK Schmidt telescopes.
The objective was to visualise any association with optical \ion{H}{ii} regions or main sequence OB stars after identification
with the {\it SIMBAD} catalogues on {\it CDS}.

\subsection{Mid- and far- infrared image archives}

The mid-infrared (mid-IR) and far-infrared (far-IR) images were
obtained from the Midcourse Space Experiment (MSX) and from the
InfraRed Astronomical Satellite (IRAS) data archives. The values
for the mid-IR fluxes were estimated from the MSX images using the
KARMA\footnote{See:
http://www.atnf.csiro.au/computing/software/karma/} package, by
estimating the flux in the source and also in the background. The
background residual flux was measured by taking the averaged flux
in boxes at diverse positions around the mm source. This method
allows removal of the background noise as well as emission from
nearby objects and extended regions. To avoid confusion with
silicate emission/absorption at 8.7~${\mu}$m (in the MSX A-band)
and with the more extended PAH contribution at 7.7, 8.6 and
11.3~${\mu}$m (in the A- and C-bands), the emission contours in
the E-band (21.3 $\mu$m) were used to estimate the extent of the
thermal dust emission. The source flux then had the background
subtracted, and the value was converted from W~m$^{-2}$~sr$^{-1}$
to Jy. The image sensitivity varies from 1.9 to $29.8{\times}10^{-7}$~W~m$^{-2}$~sr$^{-1}$, 
with an angular resolution of $18''$ (i.e. 0.18 pc at 2 kpc). The positions in the
MSX point source catalogue are accurate to within $1''-5''$.  The
conversion factors were 6.84$\times$10$^3$, 2.74$\times$10$^4$,
3.08$\times$10$^4$ and 2.37$\times$10$^4$ Jy per
W~m$^{-2}$~sr$^{-1}$ for the A-band (8.3 $\mu$m), C-band (12.1
$\mu$m), D-band (14.6 $\mu$m) and  E-band (21.3 $\mu$m),
respectively (see Egan et al. \cite{egan99} for a MSX explanatory
guide). These factors include a correction turning the 6-arcsec
square pixel into a Gaussian area. The far-IR fluxes were taken
directly from the IRAS Point Source Catalogue (IRAS PSC).

\subsection{Submillimetre and millimetre continuum observations}

The submillimetre observations at 450 and 850 $\mu$m were conducted using the Sub-mm Common User Bolometer Array (SCUBA) on the James
Clerk Maxwell Telescope (JCMT) in September 2002 for three of the five sources. G\,31.28+0.06 was observed by Walsh et al. (\cite{walsh03}). 
The observations were made in jiggle map mode with a map size $\sim$2-3 arcmin. The pixel size was
3 arcsec. Pointing and calibration were made on HL Tau. The average zenith atmospheric optical depths were 1.8 and 0.3 at 450 and 850 $\mu$m,
respectively. The volt-to-Jy-beam$^{-1}$ conversion factors were 373.1 and 260.0. The image sensitivities were $\sim0.3$ and 0.01~Jy~beam$^{-1}$
at 450 and 850 $\mu$m, respectively. The {\it FWHM} beam sizes were estimated to be 8 and 15 arcsec (i.e. 0.08-0.15 pc at 2 kpc) at 450 and 850 $\mu$m,
respectively. The data were reduced with the standard procedure SURF (reduce-switch, flatfield, extinction, remsky and
rebin). The flux density values were estimated in each image with the KVIEW procedure described in Sect. 2.2.

The 1.2-mm observations were made using the SEST IMaging Bolometer Array (SIMBA) on the Swedish-ESO Submillimetre Telescope (SEST) 
in October 2001 and June 2002.
SIMBA is a 37-channel hexagonal array in which the {\it HPBW} of a single element is about $24''$ (i.e. 0.24 pc at 2 kpc) and  the separation between
elements on the sky is $44''$. The bandwidth in each channel is about 50\,GHz. Spectral line emission present in the band
may contribute to the total continuum flux values reported in this paper, though not more than 30\%. Free-free emission is not expected to
contribute to the (sub-)mm flux given the radio-quiet nature of the targets. Assuming optically thin free-free emission for frequencies $\ge8$~GHz,
80~mJy at 8.6~GHz would convert to 57 mJy at 250~GHz, i.e. $\sim1\%$ of the 1.2-mm flux measured at the same position. Similarly, for an ionized wind
model with $\nu^{0.6}$ (Panagia \& Felli \cite{panagia75}), the extrapolation at 250~GHz of the radio fluxes would yield less than 11\% of the
observed flux densities. These observations were made in fast-mapping mode. Typical areas of $600''{\times}384''$ were imaged in order
to detect nearby IRAS sources in the field, using
a scan speed of $80''$~s$^{-1}$. The resulting pixel size is 8 arcsec. Typically, total observation times were around 15 minutes per source, giving a
sensitivity $\sim0.1$~Jy~beam$^{-1}$.
Zenith opacities ranged from 0.2 -- 0.4 throughout the observing runs in 2001 and 2002. The resulting data were reduced
with the MOPSI mapping software package developed by R. Zylka, using the ``deconvolution'' algorithm\footnote{Developed within the SIMBA
collaboration.} to remove the contribution of the
electronics arising from the fast-mapping observing mode, the ``converting'' algorithm (Salter \cite{salter83}) to convert the coordinates from
rectangular to equatorial, and partly the NOD2 and GILDAS libraries. The calibration of our data was performed using Uranus.
The multiplication factor between counts and Jy~beam$^{-1}$ was 138.8~mJy~count$^{-1}$~beam$^{-1}$ for the October 2001 run
(G\,188.95+0.89, G\,192.60-0.05) and 66.9~mJy~count$^{-1}$~beam$^{-1}$ for the June 2002 run (G31.28+0.06, G59.78+0.06). The flux density
values were estimated with the KVIEW procedure described in Sect. 2.2. The 1.2-mm fluxes for G\,173.49+2.42 were taken from Beuther et al.
(\cite{beuther02}).

\subsection{Molecular spectral line observations}

The molecular line observations were carried out in May 1999 with the Onsala-20m millimetre telescope (OSO-20m) and in October
2002 with the
Mopra millimetre telescope. CH$_3$CN (92~GHz) and C$^{18}$O (109~GHz) lines were observed toward each methanol maser position
with a single pointing. 80~MHz and 64~MHz bandwidths were used at Onsala and Mopra, respectively. The gain of the Onsala-20m telescope
is $\sim20$~Jy~K$^{-1}$ in the range 85-112 GHz while the gain of the Mopra telescope is about 30~Jy~K$^{-1}$. CH$_3$CN observations were
performed with Mopra only. C$^{18}$O observations were carried out with Mopra toward G\,192.60-0.05 and with the OSO-20m toward the
other four sources. The pointing errors were estimated to be less than 5 arcsec for the OSO-20m and less than 10 arcsec for Mopra.
The data were processed with XS, the Onsala data reduction software.

\subsection{Radio continuum data archives}

Radio continuum information was obtained from {\it Aladin} and
{\it SIMBAD} on {\it CDS} within a $4'\times4'$ region around each
methanol maser site. Specific radio survey catalogues were also
searched using {\it VizieR}. These included the 1.4-GHz NRAO VLA
sky survey (NVSS, Condon et al. \cite{condon98}), the 5-GHz VLA
survey of the Galactic plane (Becker et al. \cite{becker94}), the
survey of radio \ion{H}{ii} regions in the Northern Sky through
recombination line emission (Lockman \cite{lockman89}), the GB6
catalogue of radio sources at 4.85~GHz (Gregory et al.
\cite{gregory96}), the UC~\ion{H}{ii} region survey (Kurtz et al.
1994) and an ATCA survey at 8.6 GHz (Walsh et al. \cite{walsh98}).
Source positions from Snell \& Bally (\cite{snell86}) for
G\,192.60-0.05 and from Sridharan et al. (\cite{sridharan02}) for
G\,59.78+0.06 were also plotted.

\section{Results}

\subsection{Derivation of physical quantities}

The goal of this paper is to derive the physical characteristics of the methanol maser sites.
These physical quantities include the luminosity ($L_{total}$), the gas mass ($M_{gas}$), the gas column density
($N_{\rm H_2}$), the gas density ($n_{\rm H_2}$), the dust ($T_{cold}$ and $T_{hot}$) and gas temperatures ($T_{rot}$),
the virial mass ($M_{vir}$) and the power law index of the dust optical depth ($\beta$). The values of these physical
quantities are derived from the dust continuum emission and from the molecular line emission for each clump in the five
MSF regions studied in this paper. We do not intend to model density and temperature structures in detail. This will be undertaken
with higher angular resolution instruments in the future.

\subsubsection{Spectral Energy Distribution of dust emission}

The Spectral Energy Distribution (SED) of each millimetre source with continuum data points from 8 $\mu$m to 1.2 mm is
fitted with two grey-body functions modified with a non-constant emissivity function, which model emission from a hot ($>100$~K)
dust core (diameter, $d_{hot}$) embedded in a larger cold ($<30$~K) or warm ($30<T<100$~K) envelope (diameter, $d_{cold}$). The 
dust optical depth in the emissivity function is expressed as a power law ${\tau}_d={\tau}_0({\nu}/{\nu}_0)^{\beta}$
where ${\nu}$ is the frequency and ${\beta}$ is the dust grain emissivity index. The hot dust core is characterised by
a dust temperature ($T_{hot}$) and a solid angle ($\Omega_{hot}$) which are two free parameters in the first grey-body model.
The cold (or warm) envelope is characterised by $T_{cold}$ and $\Omega_{cold}$ in the second grey-body function. Two additional parameters,
the turn-over frequency between optically thin and thick regimes (${\nu}$=${\nu}_0$ where ${\tau}$=${\tau}_0$=1) and ${\beta}$, are also estimated.
In total six free parameters are therefore estimated using a minimum ${\chi}^2$-Levenberg-Marquardt bestfit of the functions to the data.

This model is a good approximation if each methanol maser site harbours a single young stellar object (YSO).
In the case of a multiple YSO system or a YSO cluster, our model will reproduce emission from both the individual YSOs and the cluster envelope.
However, only a few massive YSOs are expected to be present within each methanol maser site due to the relatively small region observed. The SED will
be dominated by emission from the most luminous objects present.
In summary, our model will give the SED of a single YSO at best and the total SED of a YSO
cluster at least.

The total luminosity ($L_{total}$) is derived by integrating  the SED
over the frequency/wavelength range. The gas mass is estimated using the following equation, adopting the Rayleigh-Jeans
approximation to the Planck function, and assuming optically thin emission at 1.2~mm:
\begin{equation}
M_{gas}=5\times10^{-31}\frac{F_{1.2mm}D^2{\lambda}^2}{2kT_{cold}R_d{\kappa}_d}~{\rm in~M_{\odot}},
\end{equation}
where $F_{1.2mm}$ is the 1.2-mm continuum intergrated flux in W~m$^{-2}$~Hz$^{-1}$, $D$ the distance to the source in m, ${\lambda}$
the wavelength in m, $k$ the Boltzmann's constant in J~K$^{-1}$, $T_{cold}$ the dust temperature in K, $R_d$ the dust-to-gas
mass ratio and ${\kappa}_d={\kappa}_0({\nu}/{\nu}_0)^{\beta}$ the mass coefficient absorption (${\propto}\tau_d$) in m$^2$~kg$^{-1}$.
The Rayleigh-Jeans approximation holds as long as the temperature
is greater than 40~K at 1.2~mm or 250~GHz. For temperatures less than 30~K, the use of the full Planck function is recommended.
The values of ${\kappa}_d$ and $R_d$ may vary with physical conditions.
Values of 0.1~m$^2$~kg$^{-1}$ at ${\lambda}=1.2$~mm and 1\%
are adopted for ${\kappa}_d$ and $R_d$, respectively, in the present work (Ossenkopf \& Henning \cite{ossenkopf94}). This gives values of $M_{gas}$
similar or down to 4 times less than when using the mass estimate method from Hildebrand (\cite{hildebrand83}).

$N_{\rm H_2}$ and $n_{\rm H_2}$ are calculated by dividing $M_{gas}$ by $\pi({\theta}_{sest}D/2)^2$ and $4\pi({\theta}_{sest}D/2)^3/3$,
respectively, where $\theta_{sest}\approx24''$ is the angular resolution of SEST at 1.2~mm. The values of the column density and density
in table 5 are therefore beam averaged values. $N_{\rm H_2}$ and the ${\rm H_2}$ mass density are also estimated using $d_{cold}$, which is the 
fitted diameter of the protostellar envelope (tables 5 and 6).

\subsubsection{Molecular line analysis}

The molecular lines in the CH$_3$CN and C$^{18}$O spectra are fitted with Gaussian functions. The Gaussian parameters are
given in tables 3 and 4. CH$_3$CN($5_K-4_K$) spectral lines corresponding to different $K$-components are used to derive an estimate of the
gas temperature using the rotational diagram method and assuming that CH$_3$CN emission is optically thin.
The rotation method is based on three assumptions. ({\it i}) The spectral line is optically thin. ({\it ii}) The brightness
temperature ($T_{bg}$) of the background is negligible with respect to the brightness temperature ($T_b$) of the source.
({\it iii}) The population of all observed energy levels is described by a single Boltzmann temperature $T_{rot}$.

Using the equation derived in Nummelin (1998) and the above assumptions, the rotation diagram is a plot of the left hand side
quantity in Eq. 2 versus $E_u/k$.
\begin{equation}
ln\left(\frac{8{\pi}k{\nu}^{2}_{ul}}{hc^3A_{ul}g_u}{\int}T_bdv\right)=ln\left(\frac{N}{Q(T)}\right)-\frac{E_u}{kT_{rot}},
\end{equation}
where ${\nu}_{ul}$ is the frequency of the rotational transition, $h$ is Planck's constant, c is the speed of light,
$A_{ul}$ is the absorption coefficient, $g_u$ is the degeneracy of the upper energy level, $N$ is the total CH$_3$CN column density
and $Q(T)$ is the partition function.
If the hypotheses were correct, the data points would
fall along a straight line (see Fig. 1) whose inverse of the negative slope is the
rotational temperature ($T_{rot}$). If all transitions are thermalised, the derived
temperature is an estimate of the kinematic temperature of the
molecular gas.
Note that $T_b$ is calculated from $T^{*}_{A}$, the antenna temperature
as follows:
\begin{equation}
T_b=\frac{T^{*}_{A}}{{\eta}_{mb}{\eta}_{bf}},
\end{equation}
where ${\eta}_{mb}$ is the main-beam efficiency factor and ${\eta}_{bf}$
is the beam-filling factor. In this work, we assume ${\eta}_{bf}=1$ (beam-averaged) i.e. $T_{mb}=T_b$
and ${\eta}_{mb}=0.55$ for the OSO-20m and ${\eta}_{mb}=0.7$ for Mopra.

The C$^{18}$O line is a very useful tool for estimating the H$_2$ column density. In the optically thin regime the C$^{18}$O
column density is given by Bourke et al. (\cite{bourke97}):
\begin{displaymath}
N_{\rm C^{18}O}=2.42{\times}10^{14}\frac{T_{ex}+0.88}{1-{\rm exp}(-5.27/T_{ex})}
\end{displaymath}
\begin{equation}
~~~~~~~~~~~~{\times}\frac{1}{J(T_{ex})-J(T_{bg})}\int{T_{mb}({\rm C^{18}O})dv}~{\rm in~cm^{-2}},
\end{equation}
where $J(T)=h\nu/k({\rm exp}(h\nu/kT)-1)$, $T_{ex}$ the excitation temperature, $T_{bg}$ the background temperature and
$h$ is Planck's constant.

\begin{figure}
\vspace{23.6cm}
\includegraphics{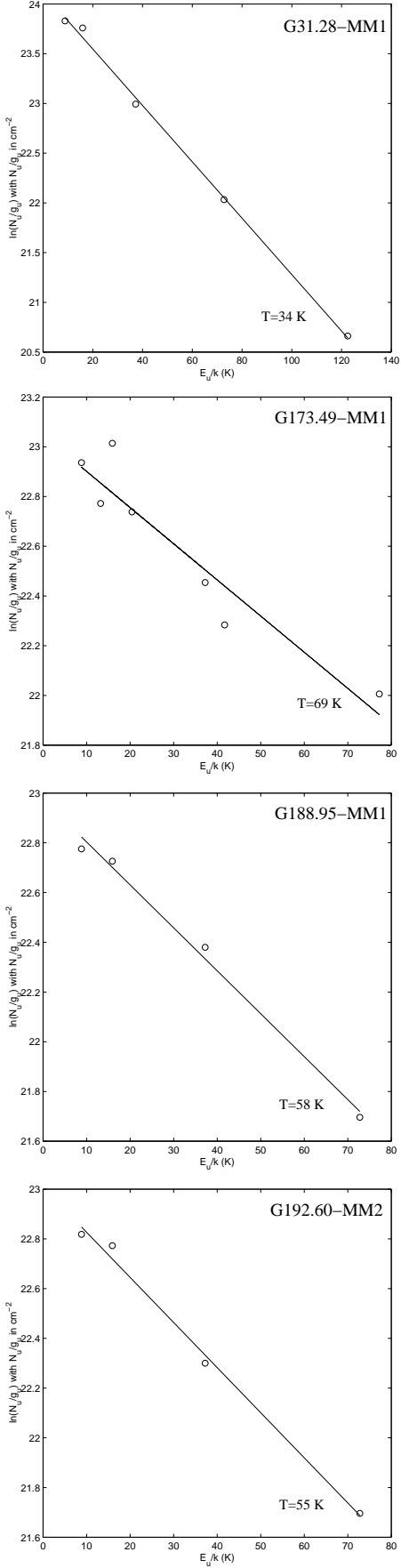}
\caption[]{Rotational diagrams where ln$(N_u/g_u)$ represents the left-hand side of Eq. 2.}
\end{figure}

Once $N_{\rm C^{18}O}$ is known, $N_{\rm H_2}$ is derived using the column density relationship in Frerking et al. (\cite{frerking82}):
\begin{equation}
N_{\rm H_2}=5.9\times10^6N_{\rm C^{18}O}+1.3\times10^{21}~{\rm in~cm^{-2}}.
\end{equation}

LTE is assumed and $T_{ex}=T_{rot}$. In this way, the rotational temperatures deduced from the CH$_3$CN data
is used to derive the C$^{18}$O column density. The deduced H$_2$ column densities are representative for the CH$_3$CN 
emitting regions if their angular sizes are similar to that of C$^{18}$O emitting regions. These assumptions will be discussed further in Sect. 4. 
The total H$_2$ mass is derived by multiplying $N_{\rm H_2}$ by $\pi(\theta_{oso}D/2)^2$, where $\theta_{oso}\approx34''$ is the
angular resolution of OSO-20m at 109.8~GHz. The density is estimated by dividing the total mass by a source volume
$V=4/3\,\pi(\theta_{oso}D/2)^3$. $M_{gas}$ and $n_{\rm H_2}$ are beam averaged values, i.e. a lower limit of the actual
values if the C$^{18}$O cloud has an angular size lower than $\theta_{oso}$.

\subsubsection{Virial mass}

The virial estimate of the molecular mass follows the approach by MacLaren et al. (\cite{maclaren88}) with density profile in $R^{-2}$, where
$R$ is the radius. The virial mass is given by:
\begin{equation}
M_{vir}=\frac{{\sigma}^2R}{G}=126{\Delta}v^2R~{\rm in~M_{\odot}},
\end{equation}
where $\sigma$ is the full 3-dimensional velocity dispersion, $R$ is the radius of the cloud in pc, ${\Delta}v$ is the FWHM of the C$^{18}$O
line in km~s$^{-1}$ and $G$ is the gravitational constant. In this work, ${\Delta}v={\Delta}v_{\rm C^{\rm 18}O}$ and $R=\theta_{oso}D/2$.
A molecular clump is gravitationally bound as long as $M_{gas}/M_{vir}>1$. Due to uncertainties in the mass estimate, the relation
$M_{gas}/M_{vir}>1$ will be discussed for each object. Note that the values of $M_{vir}$ in table 5 should be divided by 1.4 ($=34''/24''$) before
comparison with $M_{gas}$ derived from the dust continuum observations.

\begin{table*}
\caption[]{Continuum fluxes. Positions are the 1.2-mm peak
emission positions. Mid-IR fluxes are from MSX images. Sub-mm
fluxes from SCUBA images. Mm fluxes from SIMBA images. `--' means
that no emission was detected. Notes: $^*$fluxes interpolated from
SED diagrams; $^{**}$MAMBO mm fluxes from Beuther et al.
(\cite{beuther02}). References: $^1$Walsh et al. (\cite{walsh98});
$^2$Kurtz et al. (\cite{kurtz94}); $^3$Sridharan et al.
(\cite{sridharan02}); $^4$Snell \& Bally (\cite{snell86}).}
\begin{center}
\footnotesize
\begin{tabular}{lllllllllll}
\hline
\hline
    & \multicolumn{2}{c}{Position} &  \multicolumn{4}{c}{Flux(Jy) Mid-IR} &  \multicolumn{2}{c}{Sub-mm} &  Mm & Radio \\
\cline{2-3}\cline{4-7}\cline{8-9}
Source & RA     &  Dec              &  8.3      & 12.1     &  14.6     &  21.3    &   450        &    850      &  1.2 & (mJy)\\
       &(J2000) & (J2000)          &  ($\mu$m) & ($\mu$m) &  ($\mu$m) & ($\mu$m) &   ($\mu$m)   &    ($\mu$m) &  (mm) &  \\
\hline
G\,31.28-MM1  & 18 48 12.1 &  -01 26 20 &  2.4   &  3.5   &  7.5    &      34.9    &     160.0     &       19.5  &   5.4 & 80 (8.6 GHz)$^{1,2}$ \\
G\,31.28-MM2  & 18 48 10.8 &  -01 26 54 &  0.1  &  0.4   &  0.6    &       3.3    &     14.0     &        3.0  &   0.6 & $<0.4$ (15~GHz)$^2$ \\
G\,31.28-MM3  & 18 48 10.0 &  -01 27 55 &  --    &  --    &  --     &       --     &      --       &   --        &   0.3 & $<0.4$ (15~GHz)$^2$ \\
G\,59.78-MM1  & 19 43 10.0 &   23 44 59 &  --    &  --    &  --     &       --     &      --       &   --        &   0.9 & $<1$ (8.6~GHz) \\
G\,59.78-MM2  & 19 43 10.6 &   23 44 03 &  7.5   &  10.9  &  16.0   &      50.1    &      35.8$^*$ &    5.8$^*$  &   4.1 & 1 (8.6 GHz)$^3$ \\
G\,173.49-MM1 & 05 39 13.0 &   35 45 54 &  0.6   &  0.3   &  1.3    &      29.8    &      249.4    &        17.3 &   6.0$^{**}$ & $<1$ (8.6~GHz)$^3$ \\
G\,173.49-MM2 & 05 39 10.8 &   35 46 06 &  --    &  --    &  --     &       --     &       21.3    &         3.2 &   0.6$^{**}$ & $<1$ (8.6~GHz)$^3$ \\
G\,173.49-MM3 & 05 39 08.1 &   35 46 48 &  --    &  --    &  --     &       --     &       --      &         2.0 &   0.4$^{**}$ & $<1$ (8.6~GHz)$^3$ \\
G\,173.49-MM4 & 05 39 10.8 &   35 45 12 &  5.0   &   6.4  &   5.7   &      25.3    &       31.2    &         3.2 &   1.0$^{**}$ & $<1$ (8.6~GHz)$^3$ \\
G\,188.95-MM1 & 06 08 52.8 &   21 38 37 &  8.9   &  14.8  &  19.7   &      85.2    &       83.9    &        10.2 &   1.5 & $<0.4$ (15~GHz)$^2$ \\
G\,188.95-MM2 & 06 08 52.2 &   21 38 13 &  --    &  --    & 0.1$^*$ &     17.6$^*$ &      136.6    &        14.6 &   4.2 & $<0.4$ (15~GHz)$^2$\\
G\,192.60-MM1 & 06 12 52.9 &  18 00 27  &  0.2    &  0.2   &   1.0   &      17.4    &      260.0    &        26.3 &   5.4 & 22 (8.6 GHz)$^2$ \\
G\,192.60-MM2 & 06 12 53.4 &  17 59 23  &  70.6  & 126.5   &  165.4  &     301.4    &      250.0    &        28.9 &   7.4 & 3-5 (5 GHz)$^4$ \\
G\,192.60-MM3 & 06 12 56.8 &  17 58 03  &  --    &  --    &  --     &       --     &     --       &   --        &   0.8 & $<0.4$ (15~GHz)$^2$ \\
\hline
\end{tabular}
\end{center}
\end{table*}

\subsection{Presentation of the results}

For each methanol maser site, an overview of the region consisting of a (sub)mm contour map and
symbols representing radio sources, IRAS and maser sources overlaid on an optical image is given.
This is image {\bf a.} in Figs. 2-6. A close-up image of the methanol maser site is also presented
with the mid-IR and 450 $\mu$m contours overlaid on the 850 $\mu$m image. This is image {\bf b.}
in Figs. 2-6. For the mm emission sources with enough data points from mid-IR to mm wavelengths (table 1),
the Spectral Energy Distribution (SED) diagrams are presented with the physical parameters and derived physical quantities.
IRAS fluxes were used only when an association with a mm source was clearly identified (table 2).
These are plots {\bf c.} in Figs. 2-6. Finally, CH$_3$CN and C$^{18}$O spectra are presented
in Figs. 2-6~{\bf d.} with corresponding data values in tables 3 and 4. Rotational diagrams are given in Fig. 1.
Table 5 summarises the derived physical quantities for the five
methanol maser sites. To derive these values, the OSO and SEST telescope beams are mainly used as physical angular sizes because C$^{18}$O 
line emission and dust emission may not arise exactly from the same region. An estimate of the clump diameter in the SED model is 
given in Fig.~2-6.

\begin{table*}
\caption[]{IRAS fluxes and association with mm sources.}
\begin{center}
\footnotesize
\begin{tabular}{lllllllll}
\hline
\hline
Source & IRAS name & 12  & 25 & 60 & 100 & log($F_{25}/F_{12}$) & log($F_{60}/F_{12}$) & Association \\
       &           & ($\mu$m) & ($\mu$m) & ($\mu$m) & ($\mu$m) &  &  &  \\
\hline
G\,31.28+0.06  & 18456--0129 & 4.85  & 89.33 & 1071.0 & 3693.0 & 1.26 & 2.34 & G\,31.28-MM1 \\
G\,59.78+0.06  & 19410+2336 & 14.4  & 108.8 &  982.5 & 1631.0 & 0.98 & 1.83 & G\,59.78-MM2 \\
G\,173.49+2.42 & 05358+3543 & 5.6   & 74.7  & 722.3  & 1310.0 & 1.12 & 2.11  & G\,173.49-MM4 \\
G\,188.95+0.89 & 06058+2138 & 13.9  & 140.2 & 955.5  & 1666.0 & 1.00 & 1.83  & G\,188.95-MM1 \\
G\,192.60--0.05 & 06099+1800 & 107.2 & 371.6 & 3145.0 & 5285.0 & 0.54 & 1.47  & G\,192.60-MM2 \\
\hline
\end{tabular}
\end{center}
\end{table*}

\begin{table*}
\caption[]{CH$_3$CN Spectral line parameters for the methanol
maser sites derived from Gaussian fitting.}
\begin{center}
\footnotesize
\begin{tabular}{lllllll}
\hline \hline Source & $K$  & $v_{lsr}$  & $T^{*}_{A}$ & rms & ${\Delta}v$ & $\int{T^{*}_{A}}dv$  \\
       &      & (km s$^{-1}$) & (K)      & (K) & (km s$^{-1}$) & (K km s$^{-1}$) \\
\hline
G\,31.28-MM1  & 0 & 110.5 & 0.15 & 0.01 &  4.1 & 0.66 \\
            & 1 & 110.5 & 0.13 & 0.01 &  4.1 & 0.59 \\
            & 2 & 110.5 & 0.06 & 0.01 &  4.0 & 0.24 \\
            & 3 & 110.9 & 0.03 & 0.01 &  4.0 & 0.14 \\
            & 4 & 110.0 & 0.01 & 0.01 &  4.1 & 0.01 \\
G\,59.78-MM2  & 1 & 21.3 & 0.07 & 0.02 & 1.8 & 0.14   \\
            & 2 & 21.5 & 0.05 & 0.02 & 1.5 & 0.09   \\
G\,173.49-MM1 &   & see & Kalenskii   & et   & al. & (\cite{kalenskii00}) \\ 
G\,188.95-MM1 & 0 & 2.3 & 0.05 & 0.02 & 4.0 & 0.23  \\
            & 1 & 2.4 & 0.04 & 0.02 & 4.1 & 0.21  \\
            & 2 & 2.5 & 0.03 & 0.02 & 4.1 & 0.13  \\
            & 3 & 3.8 & 0.02 & 0.02 & 4.1 & 0.10  \\
G\,192.60-MM2 & 0 & 8.2 & 0.07 & 0.02 & 3.0 & 0.24  \\
            & 1 & 7.8 & 0.08 & 0.02 & 2.7 & 0.22  \\
            & 2 & 7.8 & 0.03 & 0.02 & 3.2 & 0.12  \\
            & 3 & 7.6 & 0.03 & 0.02 & 3.1 & 0.10  \\

\hline
\end{tabular}
\end{center}
\end{table*}

\begin{table*}
\caption[]{C$^{18}$O spectral line parameters for the methanol
maser sites derived from Gaussian fitting.}
\begin{center}
\footnotesize
\begin{tabular}{lllllll}
\hline \hline
Source & $v_{lsr}$ & $T^{*}_{A}$ & rms & ${\Delta}v$ & $\int{T^{*}_{A}}dv$ & ${\Delta}v_{\rm CH_3OH}$ \\
       & (km s$^{-1}$) & (K)   & (K) & (km s$^{-1}$) & (K km s$^{-1}$) & (km s$^{-1}$) \\
\hline
G\,31.28-MM1  & 108.5 & 1.89 & 0.17 & 3.6 & 7.20 & 106;114 \\
G\,59.78-MM2  &  22.5 & 1.31 & 0.07 & 2.0 & 2.83 & 16;27 \\
G\,173.49-MM1 & -16.7 & 0.51 & 0.06 & 3.2 & 1.71 & -15;-11 \\
G\,188.95-MM1 &   3.0 & 0.66 & 0.08 & 3.2 & 2.25 & 8;12 \\
G\,192.60-MM2 &   6.2 & 0.67 & 0.04 & 3.0 & 2.15 & 1;6 \\
\hline
\end{tabular}
\end{center}
\end{table*}

\subsubsection{G\,31.28+0.06}

Three mm sources are detected toward G\,31.28+0.06 (Fig. 2a). They
are distributed along a dusty filament seen with SIMBA (Fig. 2a).
The methanol maser site coexists with a \ion{H}{ii} region and
an IRAS source in the mm source G\,31.28+0.06-MM1 (hereafter G\,31.28-MM1), but is separated from the
radio source peak emission by $3.5\times10^4$~AU at a distance of 5.6 kpc. The morphology of
the \ion{H}{ii} region is irregular and clumpy, with a major
axis of about $10''$. From the values given by Kurtz et al.
(\cite{kurtz94}), the peak brightness temperatures are only
$\sim80$ K at 3.6 cm and $\sim55$ K at 2 cm. This implies that
this \ion{H}{ii} region has a low value EM, and does not belong to
the class of UC \ion{H}{ii} regions as defined by Kurtz et al.
(\cite{kurtz00}). The number of ionizing photons required to
excite the \ion{H}{ii} region is $\sim4.6{\times}10^{47}$~s$^{-1}$.
This could be supplied by an O9.5 ZAMS star, which has a
luminosity of $3.8\times10^4$~L$_{\odot}$. No OB association or optical
\ion{H}{ii} region is seen nearby to G\,31.28+0.06 although W~43
is located $25'$ south. The nature of the NVSS radio source in N-E
is unclear. G\,31.28-MM2 and -MM3 are much weaker mm emitters.
G\,31.28-MM1 and G\,31.28-MM2 also coincide with sub-mm and mid-IR
sources within the positional accuracy (Fig. 2b).
In figure 2c, a SED
model is presented for the clumps 1 and 2. The SED of G\,31.28-MM1
is consistent with that of $\sim10^5$~L$_{\odot}$ and
$\sim10^3$~M$_{\odot}$ clump. The SED of G\,31.28-MM1 is not well modelled between 8.3
and 12.1~${\mu}$m. This may be due to silicate absorption at 8.7~${\mu}$m that is not 
taken into account in our modelling. CH$_3$CN and C$^{18}$O lines (Fig. 2d) were detected toward the methanol maser site at velocities
falling in the methanol maser velocity range (106 to 114
km~s$^{-1}$). The gas column density and temperature agree quite
well with those derived from dust observations. The clump is
gravitationally bound (see table 5). G\,31.28-MM2 is less luminous
($10^3$~L$_{\odot}$), less massive ($340$ M$_{\odot}$) and much
cooler (16 K) than G\,31.28-MM1.

\subsubsection{G\,59.78+0.06}

G\,59.78+0.06 is a typical isolated maser, with no bright UC \ion{H}{ii} region associated with it (Fig. 3). The maser site, however, coincides 
with an IRAS source satisfying the Wood \& Churchwell criteria for a UC \ion{H}{ii} region, as well as a weak radio source of flux $\sim1$~mJy
(Sridharan et al. \cite{sridharan02}). The 1.2-mm continuum map reveals a ``peanut'' shape with two clumps, G\,59.78-MM1 and -MM2. G\,59.78-MM1
is the coolest
clump with no mid-IR emission. Since SCUBA observations could not be made, the SED could not be drawn. In contrast, the methanol maser site
G\,59.78-MM2 is a much warmer clump of gas and dust. IRAS data was used in compiling the SED for G\,59.78-MM2, given the close spatial coincidence
between all types of emission.
A luminosity of $\sim10^4$~L$_{\odot}$ and a mass of $\sim10^2$~M$_{\odot}$ were estimated for a distance of 2.1 kpc. Strong C$^{18}$O
emission was also detected whereas the CH$_3$CN lines are weak and thus only lower limits are given in table 5. CH$_3$CN and C$^{18}$O
lines were detected toward the methanol maser site at velocities falling in the methanol maser velocity range (16 to 27 km~s$^{-1}$).
G59.78-MM2 is gravitationally bound within the mass estimate errors.

\subsubsection{G\,173.49+2.42 (S231, S233IR)}

G\,173.49+2.42, also known as S231 in maser catalogues (e.g. Pestalozzi, Minier \& Booth \cite{pestalozzi04}) or as S233IR (e.g. Porras et al.
\cite{porras00}), is a methanol maser source without any associated visible star-forming region. The methanol maser site is surrounded by
three \ion{H}{ii} regions (Israel \& Felli \cite{israel78}),
S231 (north), S233 (north-west) and S235 (east), which are too offset to be seen in Fig.~4a. Two radio sources are observed east from the maser site
(Fig. 4a) and are associated with source S235 \#2 in Israel \& Felli (\cite{israel78}).
No radio continuum detection has been reported toward G\,173.49+2.42 in Sridharan et al. (\cite{sridharan02}). In contrast, the sub-mm
continuum map is rich in emission sites. Four objects
were detected with SCUBA: G\,173.49-MM1, G\,173.49-MM2, G\,173.49-MM3 and G\,173.49-MM4. G\,173.49-MM1 harbours the methanol maser, exhibits mid-IR emission
and coincides with one of the 1.2-mm continuum sources observed by Beuther et al. (\cite{beuther02}). G\,173.49-MM2 and G\,173.49-MM3 are only seen at
sub-mm/mm wavelengths. G\,173.49-MM4 appears to be associated with the far-IR source reported in the IRAS archives. The four mm objects are also
associated with H$_2$ nebulosities (Porras et al. \cite{porras00}). SCUBA observations of G\,173.49-MM3 are incomplete because it appears at
the edge of the map. SEDs are built for G\,173.49-MM1, G\,173.49-MM2 and G\,173.49-MM4, with only a single grey-body function fitted to the observed data
of G\,173.49-MM2. The single grey-body fit is certainly a rough estimate but it confirms that G\,173.49-MM2 is the coolest clump. G\,173.49-MM1 and G\,173.49-MM4
are modelled as massive ($>20$ M$_{\odot}$) and luminous ($>10^3$~L$_{\odot}$) clumps for a distance of 1.8 kpc. The SED of
G\,173.49-MM1 is not well modelled between 8.3
and 12.1~${\mu}$m. This may be due to silicate absorption at 8.7~${\mu}$m that is not taken into account in our modelling. CH$_3$CN and C$^{18}$O
lines were detected toward the methanol maser site (see tables 3 and 4) at velocities slightly blueshifted with respect to the methanol maser velocity
range (-15 to -11 km~s$^{-1}$). Column densities toward G\,173.49-MM1 derived from gas and dust observations agree
within a factor of 2 (see table 5).  G\,173.49-MM1 is gravitationally bound within the mass estimate errors.

\subsubsection{G\,188.95+0.89 (S252, AFGL\,5180)}

G\,188.95+0.89 is a methanol maser site seen in projection nearby Sh\,2-247, a complex of \ion{H}{ii} regions and OB stars (fig. 5a).
Another radio source (G\,188.95+0.91) to the east is an unknown and unresolved nebula (Kurtz et al. \cite{kurtz94}). No radio source has
been reported near the location of the maser site. 1.2-mm continuum
emission is detected toward the methanol maser position. The mm map is
divided into two clumps, G\,188.95-MM1 and -MM2. At a distance of 2.2 kpc they are just resolved
with the beam of SIMBA/SEST. G\,188.95-MM1 is the methanol maser site and exhibits emission from mid-IR to mm.
G\,188.95-MM2 is also a strong sub-mm and mm source, but does not coincide with any mid-IR peak emission. A third region to the south-east
appears in the MSX contour map. The far-IR emission observed with IRAS probably includes emission from both G\,188.95-MM1 and -MM2. 
The SED of G\,188.95-MM1 requires a  large mass ($50$~M$_{\odot}$) and high luminosity
($\sim1.1\times10^4$~L$_{\odot}$). The SED of G\,188.95-MM1 is not well modelled between 8.3 and 12.1~${\mu}$m. This may be due to silicate 
absorption at 8.7~${\mu}$m that is not taken into account in our modelling.
Similarly, the SED of G\,188.95-MM2, modelled with a single grey-body function, also requires a high mass and luminosity. Note that a total mass of
60~M$_{\odot}$ is given in Lada \& Lada (\cite{lada03}) for AFGL~5180 at a distance of 1.5~kpc. 170~M$_{\odot}$ in total are derived from SIMBA
observations
for G\,188.95+0.89 at a distance of 2.2~kpc. By taking 1.5 kpc instead of 2.2 kpc in eq. 1, a total mass of 78~M$_{\odot}$ would have been 
calculated.
Our result is consistent with the Lada \& Lada's mass estimate. CH$_3$CN and C$^{18}$O lines were detected toward the methanol maser site
(see tables 3 and 4) at velocities blueshifted with respect to the methanol maser velocity range (8 to 12 km~s$^{-1}$).
Similar column densities are derived from gas and dust emission. The gravitational status of G\,188.95-MM1 is unclear.

\subsubsection{G\,192.60-0.05 (S255IR)}

G\,192.60-0.05 is a methanol maser site situated between two \ion{H}{ii} regions, S255 and S257 (see Minier et al. \cite{minier01} and 
references therein), consisting of OB stars (fig. 6a). SIMBA observations at 1.2 mm reveal a dusty filament elongated north-south, centred 
at the location of the methanol maser. Three mm sources are identified: G\,192.60-MM1, G\,192.60-MM2 and G\,192.60-MM3. An additional mm 
core might be present at a median position between G\,192.60-MM1 and G\,192.60-MM2 (Fig. 6b), but is too poorly resolved to be successfully 
fitted as a compact source. G\,192.60-MM1 is also concident with a UC \ion{H}{ii} region, a sub-mm source, and a mid-IR source (fig. 6b). 
G\,192.60-MM2 is the site of the methanol maser. It also contains three weak and very confined radio sources that might be interpreted as 
hyper-compact \ion{H}{ii} regions (Minier et al. \cite{minier01} for explanation). It is also associated with mid-IR, far-IR and sub-mm emission.
The IRAS source coincides with the maser site. G\,192.60-MM3 is a colder source that is not clearly identified as a clump. G\,192.60-MM1 
and G\,192.60-MM2 are modelled as two spherical clumps with a high mass ($>200$ M$_{\odot}$) and luminosity ($>10^4$~L$_{\odot}$) for 
a distance of 2.6 kpc.
The SED of G\,192.60-MM1 is not well modelled between 8.3 and 12.1~${\mu}$m. This may be due to silicate absorption at 8.7~${\mu}$m.
CH$_3$CN  and C$^{18}$O lines were detected toward the methanol maser site (see tables 3 and 4) at velocities redshifted with respect to the methanol
maser velocity range (1 to 6 km~s$^{-1}$). Column densities in  G\,192.60-MM2 derived from gas
and dust observations agree within a factor 5 (see table 5). G\,192.60-MM2 is gravitationally bound within the mass estimate errors.

\subsection{Dust grain emissivity index ($\beta$)}

The dust grain emissivity index $\beta$ varies between 1.6 and 1.9 for all sources modelled, with the exception of G\,59.78-MM2, whose SED is
poorly constrained with the absence of sub-millimetre data. This range of values is
consistent with the numerical estimates of $\beta$  presented by Ossenkopf \& Henning (\cite{ossenkopf94}), who found values of $\beta$ ranging between 1.8 and 2
for dust grains with ice mantles. The adopted value of ${\kappa}_d=0.1$~m$^2$~kg$^{-1}$ at ${\lambda}=1.2$~mm taken from Ossenkopf \& Henning
(\cite{ossenkopf94}) is therefore self-consistent.

\begin{table*}
\caption[]{Physical properties of the methanol maser sites derived from the SED and spectral line analysis. $N_{\rm H_2}$ and $n_{\rm H_2}$ are
derived from $M_{gas}$ in the continuum case by using an angular size of $24''$. The values of $N_{\rm H_2}$ derived using $d_{cold}$
are given in parentheses. $M_{gas}$ in the spectral line case is derived from $N_{\rm H_2}$ by using an angular size of $34''$.}
\begin{center}
\footnotesize
\begin{tabular}{llllllllllll}
\hline
\hline
     & \multicolumn{5}{c}{Continuum} & & \multicolumn{5}{c}{Spectral line}  \\
\cline{2-6}\cline{8-12}
Source & $T_{cold}$ &  $M_{gas}$ & $N_{\rm H_2}$ & $n_{\rm H_2}$ & $L_{total}$ & & $T_{rot}$ & $M_{gas}$ & $N_{\rm H_2}$ &  $n_{\rm H_2}$ & $M_{vir}$ \\
 & (K) & ($10^2{\rm M}_{\odot}$) & ($10^{22}$ & ($10^4$ & ($10^4$L$_{\odot}$) & & (K) & ($10^2{\rm M}_{\odot}$) & ($10^{22}$ & ($10^4$ & ($10^2{\rm M}_{\odot}$) \\
 &     &      &  ${\rm cm}^{-2}$) & ${\rm cm}^{-3}$) &  &  &  &  & ${\rm cm}^{-2}$) &  ${\rm cm}^{-3}$) &  \\
\hline
G\,31.28-MM1  & 39 & 12.6 & 17.5 (90.3) & 13.1 & 10.3 & & 34   & 21.6 & 15.1 & 8.0  & 7.4   \\
G\,59.78-MM2  & 53 & 1.0  & 9.8 (36.5) & 19.5  & 1.1  & & $>18$ & $>0.8$  & $>3.8$  & $>5.4$  & 0.9   \\
G\,173.49-MM1 & 49 & 1.2  & 15.5 (31.8) & 36.0 & 4.9  & & 69   & 1.0  & 6.5  & 10.8 & 1.8   \\
G\,188.95-MM1 & 42 & 0.5  & 4.5 (41.0) & 8.6  & 1.1  & & 58   & 1.6  & 7.4  & 10.0 & 2.3   \\
G\,192.60-MM2 & 42 & 3.2 & 22.3 (59.5) & 37.3 & 5.1 &  & 55   & 1.5  & 5.3  & 6.3  & 2.4  \\
\hline
\end{tabular}
\end{center}
\end{table*}

\begin{figure*}
\vspace{21.5cm}
\includegraphics{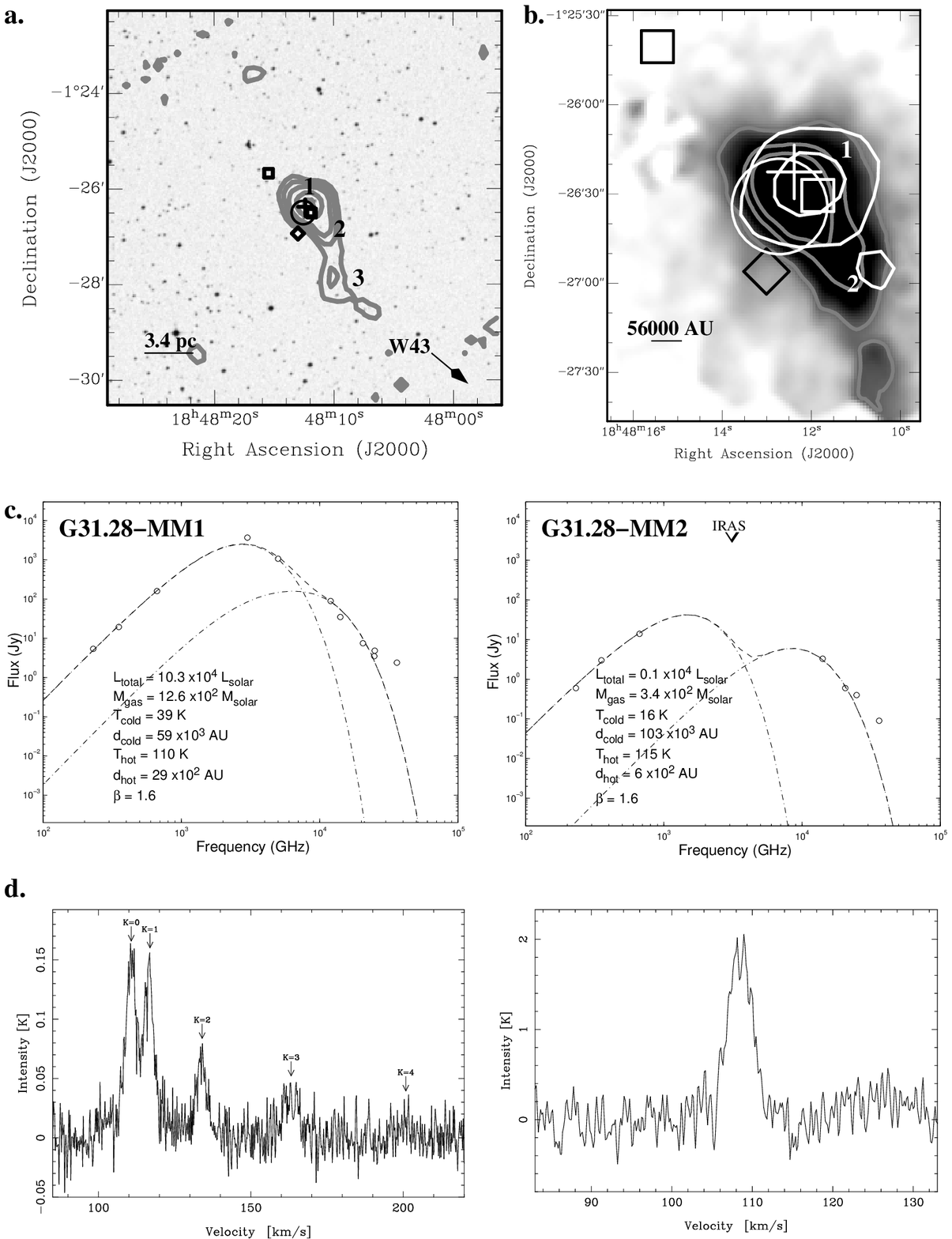}
\caption[]{G\,31.28+0.06. {\bf a.} Optical image (grey scale) of the neighbourhood
of G\,31.28+0.06. The grey contours represent the 1.2-mm continuum emission imaged by SIMBA (5, 10, 20, 50 and 90\% of peak flux).
The cross represents the position of the methanol masers, the squares are the radio continuum sites, the diamond is a recombination line
site and the large circle shows the
position of the IRAS source. The methanol maser source is not coincident with any visible object or optically visible
H\mbox{\sc ii} region but it is associated with a strong mm source. {\bf b.} Close-up of the mm sources 1 and 2.
850-$\mu$m SCUBA (grey scale), 450-$\mu$m SCUBA (grey contours and 10, 50 and 90\% of peak flux) and 21-$\mu$m MSX (white
contours and 20 and 60\% of peak flux) images. The mm sources 1 and 2 are also
seen with SCUBA and MSX. {\bf c.} SEDs of the mm sources 1 and 2. The IRAS icon shows the IRAS 100-$\mu$m flux level.
{\bf d.} CH$_3$CN and C$^{18}$O line spectra.}
\end{figure*}

\begin{figure*}
\vspace{21.5cm}
\includegraphics{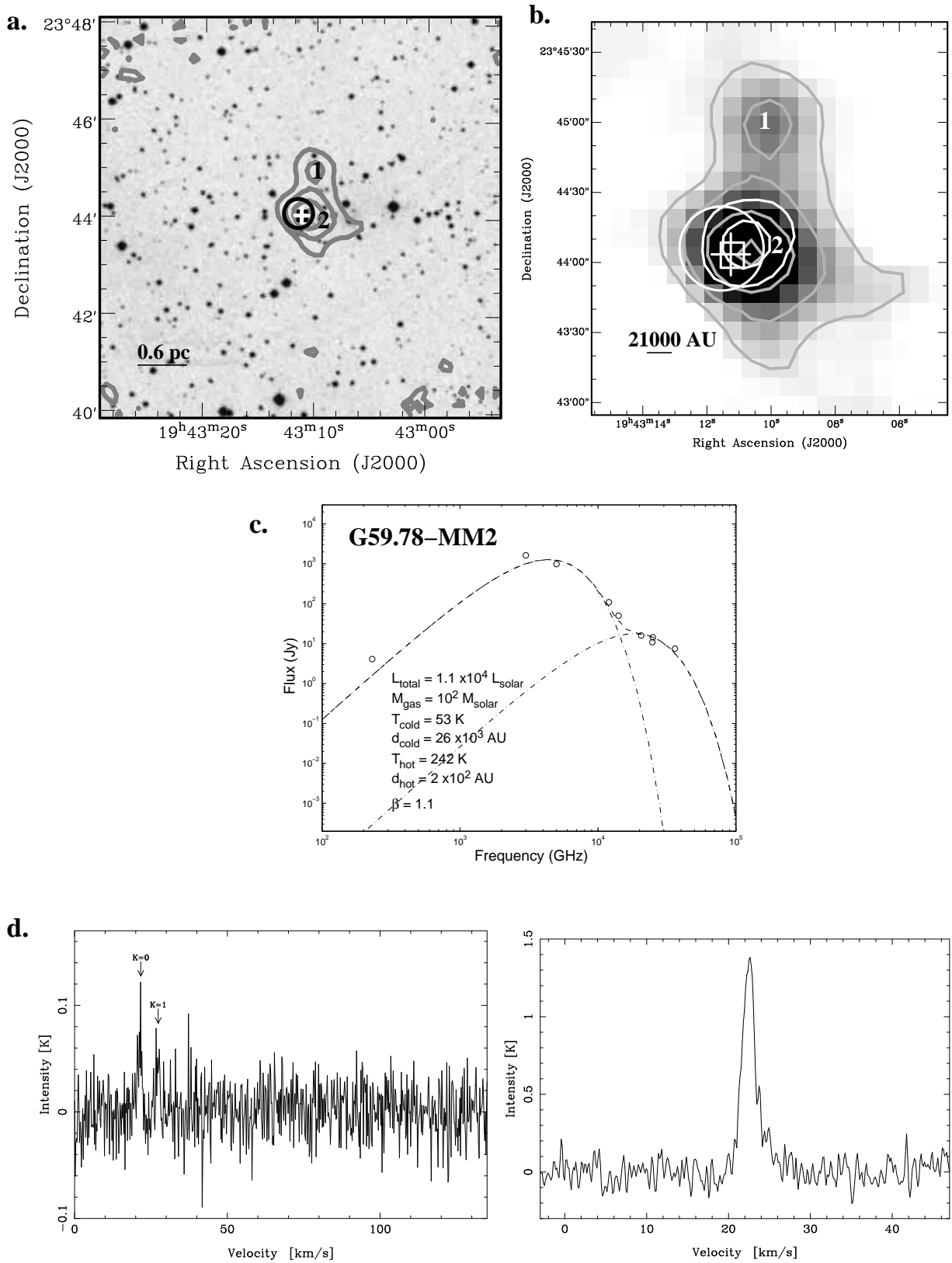}
\caption[]{G\,59.78+0.06. {\bf a.} Optical image (grey scale) of the neighbourhood
of G\,59.78+0.06. The grey contours represent the 1.2-mm continuum emission imaged by SIMBA (10, 20, 50 and 90\% of peak flux).
The cross represents the position of the methanol masers, the squares are the radio continuum sites and the large circle shows the
position of the IRAS source. The methanol maser source is not coincident with any visible object or optically visible
H\mbox{\sc ii} region but it is associated with a strong mm source. {\bf b.} Close-up of the mm sources 1 and 2.
1.2-mm SIMBA (grey scale and contours; 10, 20, 50 and 90\% of peak flux) and 21-$\mu$m MSX (white
contours; 20 and 60\% of peak flux) images. The mm source 2 is also seen with MSX. {\bf c.} SED of the mm source
2. {\bf d.} CH$_3$CN and C$^{18}$O line spectra.}
\end{figure*}

\begin{figure*}
\vspace{21.5cm}
\includegraphics{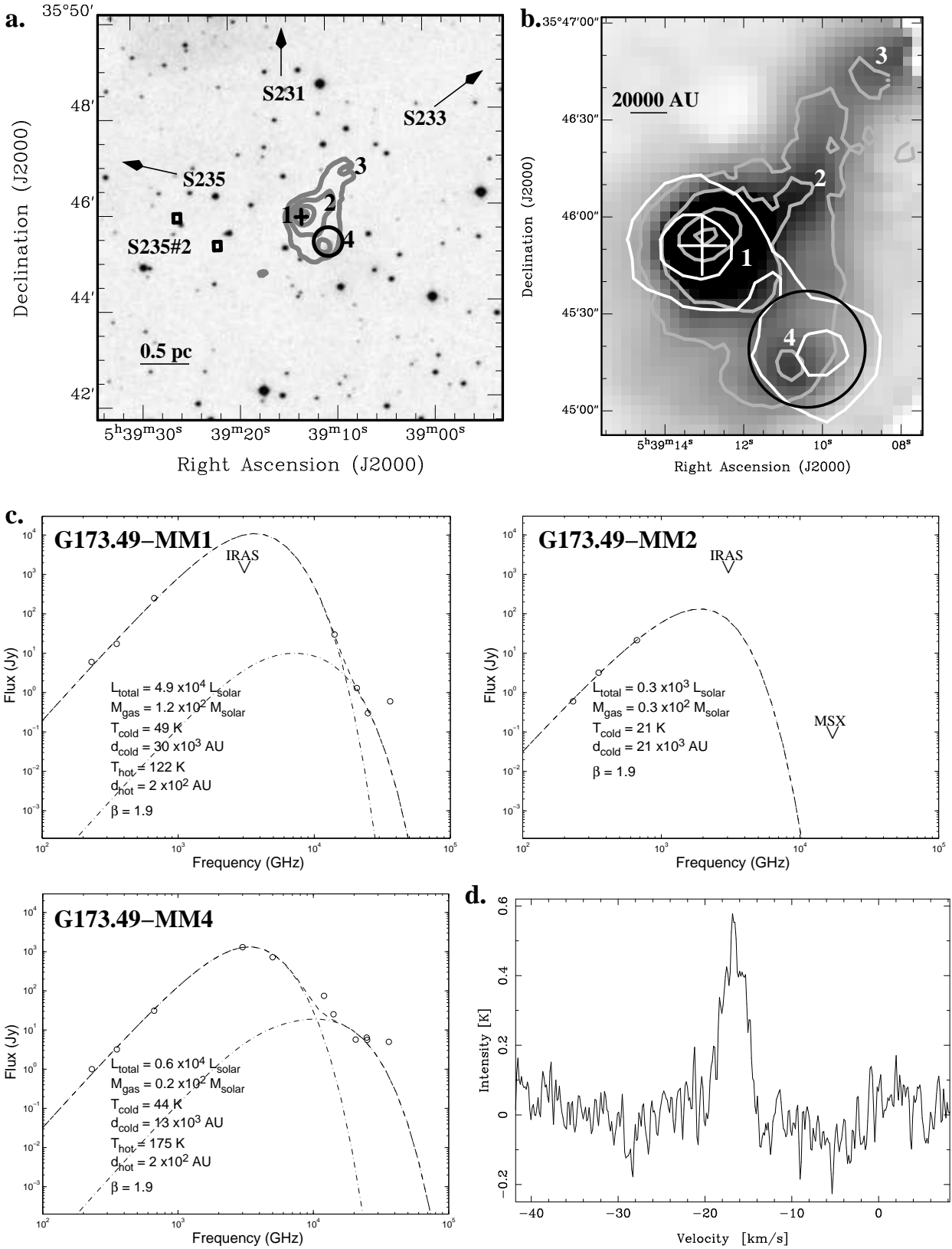}
\caption[]{G\,173.49+2.42 (S231). {\bf a.} Optical image (grey scale) of the neighbourhood
of G\,173.49+2.42. The grey contours represent the  850-$\mu$m continuum emission imaged by SCUBA (10, 20, 50 and 90\% of peak flux). Note
that the SCUBA map does not entirely cover the optical map.
The cross represents the position of the methanol masers, the squares are the radio continuum sites and the large circle shows the
position of the IRAS source. The methanol maser source is not coincident with any visible object or optically visible
H\mbox{\sc ii} region but it is associated with a strong mm source. {\bf b.} Close-up of the mm sources 1, 2, 3 and 4.
850-$\mu$m SCUBA (grey scale), 450-$\mu$m SCUBA (grey contours; 10, 20, 50 and 90\% of peak flux) and 21-$\mu$m MSX (white
contours; 20 and 60\% of peak flux) images. The mm sources are also seen with SCUBA, and MM1 and MM4 are seen with MSX.
{\bf c.} SEDs of the mm sources 1, 2 and 4. The IRAS and MSX icons show the IRAS 100-$\mu$m flux level and MSX 20-$\mu$m image noise
rms level. {\bf d.} C$^{18}$O line spectrum.}
\end{figure*}

\begin{figure*}
\vspace{21.5cm}
\includegraphics{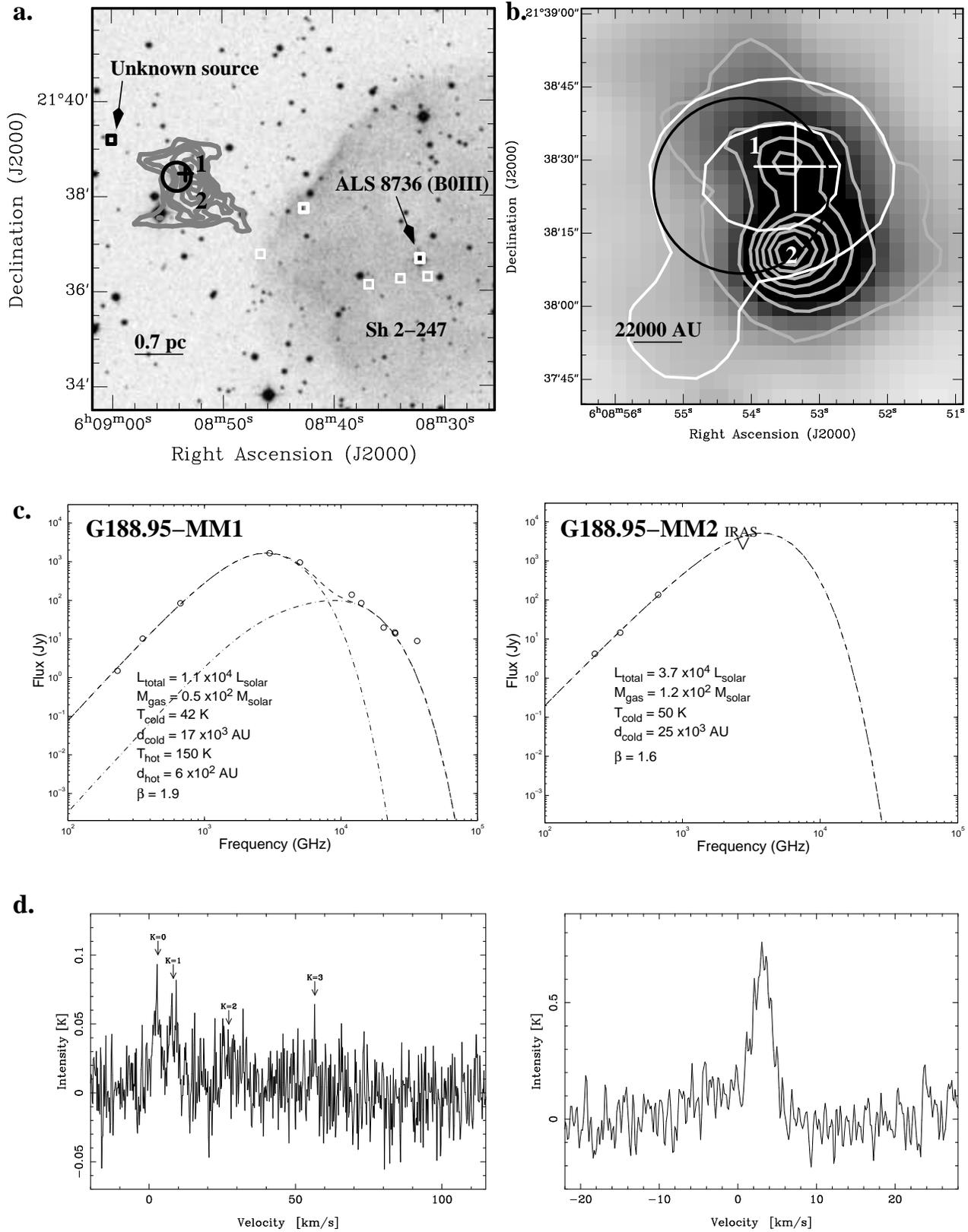}
\caption[]{G\,188.95+0.89 (AFGL\,5180). {\bf a.} Optical image (grey scale)
of the neighbourhood
of G\,188.95+0.89. The grey contours represent the 1.2-mm continuum emission imaged by SIMBA (10, 20, 30, 50. 70 and 90\% of peak
flux). The cross represents the position of the methanol masers, the squares are the radio continuum sites and the large circle
shows the position of the IRAS source. The methanol maser source is not coincident with any visible object or optically visible
H\mbox{\sc ii} region but it is associated with a strong mm source. {\bf b.} Close-up of the mm sources 1 and 2.
850-$\mu$m SCUBA (grey scale), 450-$\mu$m SCUBA (grey contours; 30, 40, 50, 60, 70 80 and 90\% of peak flux) and 21-$\mu$m MSX (white
contours; 20 and 60\% of peak flux) images. The mm sources 1 and 2 are also seen with SCUBA and MSX. {\bf c.} SEDs of mm sources
1 and 2. The IRAS icon shows the IRAS 100-$\mu$m flux level. {\bf d.} CH$_3$CN and C$^{18}$O line spectra.}
\end{figure*}

\begin{figure*}
\vspace{21.5cm}
\includegraphics{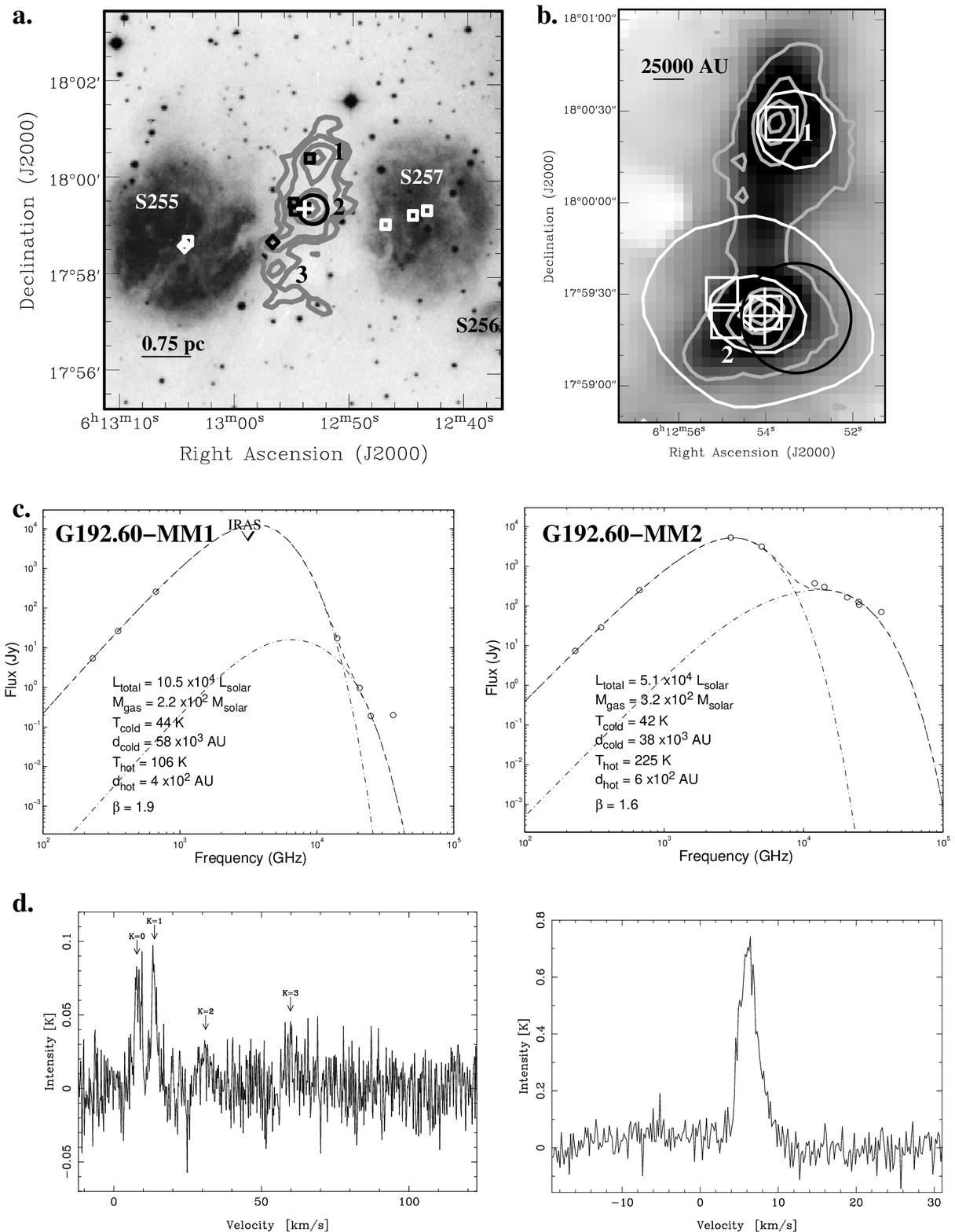}
\caption[]{G\,192.60-0.05 (S255IR). {\bf a.} Optical image (grey scale) of the neighbourhood
of G\,192.60-0.05. The grey contours represent the 1.2-mm continuum emission imaged by SIMBA (10, 20, 50 and 90\% of peak flux).
The cross represents the position of the methanol masers, the squares are the radio continuum sites, the diamond is a recombination line
site and the large circle shows the position of the IRAS source. The methanol maser source is not coincident with any visible object or
optically visible
H\mbox{\sc ii} region but it is associated with a strong mm source. {\bf b.} Close-up of the mm sources 1 and 2.
850-$\mu$m SCUBA (grey scale), 450-$\mu$m SCUBA (grey contours; 25, 45, 65 and 85\% of peak flux) and 21-$\mu$m MSX (white
contours; 5, 45 and 85\% of peak flux) images. The mm sources 1 and 2 are also seen with SCUBA and MSX.
{\bf c.} SEDs of mm sources 1 and 2. The IRAS icon shows the IRAS 100-$\mu$m flux level. {\bf d.} CH$_3$CN and C$^{18}$O line spectra.}
\end{figure*}

\begin{figure}
\vspace{20cm}
\includegraphics{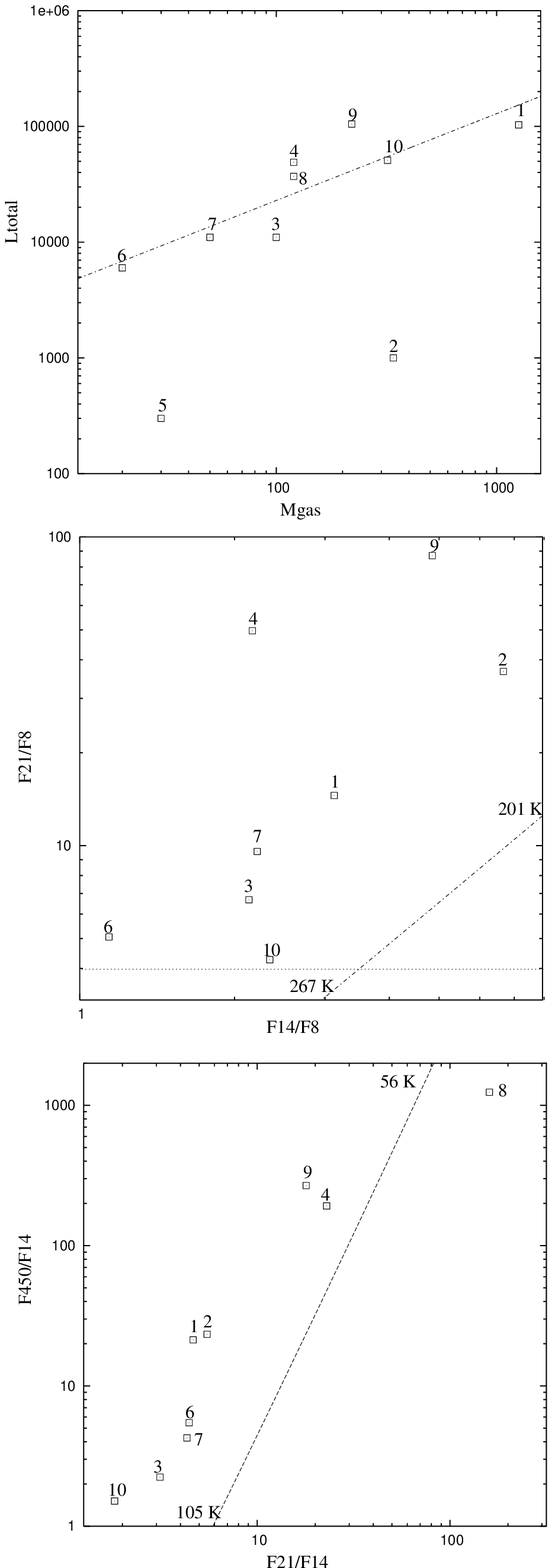}
\caption[]{Luminosity versus mass and colour diagrams. Each number represents an object as follows
G\,31.28-MM1(\#1), G\,31.28-MM2(\#2), G\,59.78-MM2(\#3), G\,173.49-MM1(\#4), G\,173.49-MM2(\#5), G\,173.49-MM4(\#6), G\,188.95-MM1(\#7),
G\,188.95-MM2(\#8), G\,192.60-MM1(\#9), G\,192.60-MM2(\#10). The dashed line in the top plot represents $L_{total}{\propto}M_{gas}^{0.75}$.
The dashed lines in the colour diagrams represent a black body  function with the temperature ranges indicated on the plots. The
dotted line in the middle plot represents the massive YSO colour criterion from Lumsden et al. (\cite{lumsden02}).}
\end{figure}

\section{Discussion}

\subsection{Methanol masers and MSF}

The aim of this work was to study the large-scale ($\sim0.5-5$ pc) environment of methanol masers. The results demonstrate that each
radio-quiet maser site is always associated with a massive ($>50$ M$_{\odot}$) and very luminous ($>10^4$ L$_{\odot}$) clump of
molecular gas and dust. Whether the clumps are very young clusters, i.e. {\it protoclusters} ($\le$ a few $10^5$ yr), forming high-mass stars
or single protostellar cores is the subject of this discussion.

These individual clumps are at best partially resolved with the relatively low angular resolution of SIMBA, SCUBA or MSX.
An estimate of their diameter is given by $d_{cold}$ in Fig. 2-6. It varies between $17\times10^3$ and $59\times10^3$~AU, i.e. $\sim$0.1 to
0.3~pc. This implies H$_2$ mass densities of $6\times10^4$ to $10^5$ M$_{\odot}$ pc$^{-3}$ in the clumps with a methanol maser (table 6). 
However, the
uncertainties on the clump diameters is of crucial importance as the density value varies with $d_{cold}^3$. By taking the
beam size of $\sim24''$ as an upper limit for each clump diameter, the mass density would be $5\times10^3$ to $3\times10^4$ M$_{\odot}$ pc$^{-3}$
at heliocentric distances of 1.8 to 5.6 kpc. These values agree with those determined for star-forming clusters such as W3:IRS5
(Megeath et al. \cite{megeath96}),
Mon~R2 (Carpenter et al. \cite{carpenter97}), NGC~2024 in Orion B and the Trapezium (Lada \& Lada \cite{lada03}). More generally, our results
are consistent with the physical parameters describing the least evolved embedded clusters in Lada \& Lada (\cite{lada03}). Embedded clusters
are all associated with massive (100-1000 M$_{\odot}$) and dense ($n_{\rm H_2}=10^4-10^5$~cm$^{-3}$) clumps with diameters $\sim0.5-1$ pc and
mass density $\sim10^3-10^4$~M$_{\odot}$ pc$^{-3}$ within the molecular clouds. Interestingly, the
overall mass, luminosity and linear size for each star-forming complex are not much larger than those measured 
for protoclusters forming stars less massive
than B2 stars such as L1688 in $\rho$-Oph with $L_{total}=6.7\times10^3$~L$_{\odot}$ (due to HD\,147889, a B2 star as well as the nearby Sco
OB2 association), $M\sim500$~M$_{\odot}$ and a diameter $\sim1$~pc (Greene \& Young \cite{greene89}; Wilking \& Lada \cite{wilking83}).

The visual extinction of the clumps with a methanol maser can be estimated with $A_v=N_{\rm H_2}/0.94\times10^{21}$~mag (Frerking et al.
\cite{frerking82}). It varies from 40 to 240 mag, implying a high degree of embeddedness and presumably a very young age ($\ll$ $10^6$ yr).
In comparison, the visual extinction in the $\rho$-Oph central C$^{18}$O cloud in L1688 is in the range 50-100 mag
(Wilking \& Lada \cite{wilking83}) with a column density $\sim10^{23}$~cm$^{-2}$ and a diameter $\sim0.5$~pc. Three mm cores (Oph-C, Oph-CS
and Oph-E) were mapped by Motte et al. (\cite{motte98}) in the central cloud and have column densities ranging from $\sim7\times10^{22}$ to
$10^{23}$~cm$^{-2}$, which compare to those in the methanol maser clumps ($4.5\times10^{22}$ to $2.2\times10^{23}$~cm$^{-2}$).

In summary, the methanol maser sites exhibit the properties of young, deeply embedded clusters of YSOs. Only luminosities in excess of
$10^4$~L$_{\odot}$ ($\sim$ZAMS B0.5, Panagia \cite{panagia73}) for clumps less than $\sim1$~pc in diameter clearly suggest the
presence of high-mass ($>8$~M$_{\odot}$) YSOs.

The detection of CH$_3$CN and C$^{18}$O toward each methanol maser site indicates that they are within HMCs with gas density
$\geq10^5$~cm$^{-3}$. CH$_3$CN emission usually arises from the 0.1-pc inner part of the hot core although ground state CH$_3$CN
emission could trace a cooler region in HMCs than those probed with vibrationally excited CH$_3$CN lines (Olmi et al. \cite{olmi03}). 
This might explain the relatively low rotational temperatures (18-69~K) in table 5. The C$^{18}$O antenna temperatures (table 4) are
much larger than those measured for CH$_3$CN (table 3). A possible explanation is that C$^{18}$O emission arises from a larger area than CH$_3$CN
emission. The CH$_3$CN rotational temperatures agree well with the $T_{cold}$ values (table 5), which suggest that CH$_3$CN($5_K-4_K$) and 
dust emission 
arises from the same area. The fitted source solid angles ($\Omega_{cold}$ and $\Omega_{hot}$ used to estimate $d_{cold}$ and $d_{hot}$) in 
the dust emission models are all smaller than the OSO and Mopra telescope beam sizes, which then probably overestimate the true size of the 
emitting regions. In consequence, the densities ($\sim10^5$~cm$^{-3}$) derived from the C$^{18}$O data (table 5) with the telescope beam sizes as 
the source angular sizes are lower limits of the CH$_3$CN clump densities.  
In addition to CH$_3$CN emission, NH$_3$ and CH$_3$OH quasi-thermal lines were 
also detected toward G\,59.78+0.06, G\,173.49+2.42, G\,188.95+0.89 and G\,192.60-0.05 (Minier \& Booth \cite{minier02}). The presence of the 
typical HMC species (CH$_3$CN, NH$_3$ and CH$_3$OH) suggest again a very young age ($\le$ a few $10^5$ yr) for the methanol maser sites 
(e.g. Rodgers \& Charnley \cite{rodgers01}).

In conclusion, methanol masers trace very luminous, deeply embedded, massive, molecular clumps in the five star-forming regions considered here.
In a few of the clumps (G\,31.28-MM1, G\,192.60-MM2 and G\,59.78-MM2), ionised sources were also detected in the radio. This implies that 
high-mass stars are currently forming within these clumps and that the deeply embedded star-forming clumps are protoclusters of high-mass YSOs.

Yet, whether methanol masers are themselves {\it directly} associated with a massive YSO remains unclear. In G\,31.28+0.06, the maser is offset
by $3.5\times10^4$ AU from the UC \ion{H}{ii} region (Walsh et al. \cite{walsh98}). The maser site in G\,59.78+0.06 appears to coincide with a very
weak radio source within $2''$ (Sridharan et al. \cite{sridharan02}).
In contrast, no radio continuum emission has been detected in G\,173.49+2.42 and in G\,188.95+0.89. Finally, the methanol maser in G\,192.60-0.05
coincides with a 4.4-mJy radio source. Assuming that a maser site and a radio continuum source coincident in the maps are indeed co-spatial, 
the methanol masers in G\,59.78+0.06 and G\,192.60-0.05 are associated with the precursors of a UC \ion{H}{ii} region.
In G\,31.28+0.06, G\,173.49+2.42 and G\,188.95+0.89,
further high angular resolution and high sensitivity mm continuum observations are required to identify the nature of the maser sources.

\subsection{Multiple star-forming (proto)clusters}

A secondary aspect of this work is the detection of objects only seen in mm continuum (hereafter ``mm-only'' clumps) and the variation
of the SED profiles with the evolutionary status of the YSOs.

\subsubsection{Cold cores}

G\,59.78-MM1 and G\,173.49-MM2 are typical mm-only clumps. G\,31.28-MM3,
G\,173.49-MM3 and G\,192.60-MM3 are other possible mm-only clumps, making in total five mm-only clumps. G\,188.95-MM2 probably emits in mid-IR, but
its flux was not measured due to confusion with the main mid-IR source, G\,188.95-MM1. These mm-only clumps are cold ($\sim20$~K for G\,173.49-MM2).
The modelled luminosity in G\,173.49-MM2 is only 300 L$_{\odot}$. Assuming
$T_{cold}=20$~K for G\,31.28-MM3, G\,59.78-MM1, G\,173.49-MM3, G\,192.60-MM3, masses of $\sim25-178$~M$_{\odot}$ are derived for the five mm-only clumps.
Excluding G\,31.28-MM3 at 5.6 kpc, masses range $19-103$~M$_{\odot}$ for G\,59.78-MM1, G\,173.49-MM2, G\,173.49-MM3 and G\,192.60-MM3. The modelled diameter
of G\,173.49-MM2 is only $21\times10^3$~AU, i.e. 0.1 pc. Using an angular size of $24''$ for each object, column densities
$\sim2.5-7.4\times10^{22}$~cm$^{-2}$ and visual extinction $\sim26-78$~mag are estimated. The physical parameters of all detected mm-only clumps
are summarised in table 6.

The cold clumps might be prestellar clumps or protostellar
clumps in the process of forming stars. Unfortunately, such types of clumps have marginally been detected and it is difficult to compare the mm-only
clump properties with those of statistical samples of massive prestellar and early protostellar clumps. A good protostar candidate in an early
accretion phase is IRAS~23385+6053 (Fontani et al. \cite{fontani03}), which is a cold ($\sim40$~K) and small ($\sim0.03$~pc) molecular core
embedded in a colder ($\sim15$~K) and larger ($<0.4$~pc) halo. The SED, luminosity ($150$ L$_{\odot}$) and temperature ($15$~K) of
IRAS~23385+6053 resemble those of G\,173.49-MM2, when the IRAS points are excluded (Fig.~13 in Fontani et al. \cite{fontani03}). But as stated by
the authors, it represents a lower limit. The luminosity of G\,173.49-MM2 could also be underestimated without the knowledge of the far-IR flux.

\subsubsection{SED profiles}

In addition to the mm-only clumps, three other types of SED
profile are identified. They might represent different
evolutionary phases or result from geometrical effects that modify
our perception of the YSO environments in function of the
line-of-sight. G\,31.28-MM2 is the only example of a cold (16~K)
envelope with a nearly hot (115~K) core. The envelope continuum emission
and the core continuum emission peak at relatively low
frequencies. G\,59.78-MM2, G\,173.49-MM4, G\,188.95-MM1 and G\,192.60-MM2
exhibit SEDs with a ``warm'' ($\sim45$~K) component peaking in the
far-IR at $60-100$~${\mu}$m and a hot ($150-250$~K) component
peaking in the mid-IR at $15-30$~${\mu}$m. They are all very
luminous ($>10^4$~L$_{\odot}$). 
Finally, G\,31.28-MM1, G173.49-MM1, G192.60-MM1 and perhaps G\,188.95-MM2 see their SED profile dominated
by a strong far-IR component. The core component of the SED has
shifted to longer wavelengths ($\sim50$~${\mu}$m) and cooled down
to $\sim110$~K. The warm ($\sim40$~K) envelope emission has
strengthened and still peaks around 100~${\mu}$m. 
The temperatures of all the ``warm'' components are in good agreement with 
those expected given the luminosities, sizes and $\beta$ (e.g. see eq. 11
in Garay \& Lizano \cite{garay99}). The hot component most likely probe
the central part ($d_{hot}=100$ to 3000 AU in Fig. 2-6) of the core around the luminous protostellar object.

To compare the evolutionary status of each object, arguments are
presented below and in Fig. 7. Only the clumps for which the SED
can be built are considered. G\,173.49-MM2 and G\,31.28-MM2 present
$L_{submm}/L_{total}>10^{-2}$ while the other clumps have
$L_{submm}/L_{total}<3\times10^{-3}$, where $L_{submm}$ is the luminosity 
integrated between 1.2 mm and 450 ${\mu}$m. Similar results are noticed
when considering $L_{total}/M_{gas}$ ratios, which are also
independent of distances. G\,173.49-MM2 and G\,31.28-MM2 have lower
$L_{total}/M_{gas}$ ratios (3-10 L$_{\odot}$~M$_{\odot}^{-1}$)
than those measured for the other clumps ($>80$
L$_{\odot}$~M$_{\odot}^{-1}$). This could be explained if
G\,173.49-MM2 and G\,31.28-MM2 are younger than the other objects
associated with maser sites and UC~\ion{H}{ii} regions, hence less
luminous. This is also seen in Fig.~7-top where G\,173.49-MM2 (\#5
in Fig. 7) and G\,31.28-MM2 (\#2 in Fig. 7) lie below the other
sources whose luminosities follow
$L_{total}{\propto}M_{gas}^{0.75}$. The objects 
G\,59.78-MM2, G\,173.49-MM4, G\,188.95-MM1 and G\,192.60-MM2 are all
relatively hot inside, and strongly emit in the mid-IR with
respect to the sub-mm emission (Fig.~7 bottom). They are all good
HMC candidates. 
The mid-IR component is very much absorbed in the FIR component
in the SEDs of G\,31.28-MM1, G\,173.49-MM1, G\,188.95-MM2 and
G\,192.60-MM1. G\,31.28-MM1 and G\,192.60-MM1 contain \ion{H}{ii} regions. The cases of 
G\,173.49-MM1 and G\,188.95-MM2 are unclear and they might belong to the HMC or 
UC~\ion{H}{ii} class of objects.

Alternatively, the least massive clumps near the methanol maser site could be low- or intermediate-mass star-forming protoclusters. 
Their physical characteristic
resemble those of the $\rho$-Oph mm cores (Motte et al. \cite{motte98}). Indeed G\,173.49-MM4 is coincident with S233IR NE, a cluster of low-mass
YSOs seen in near-IR (Porras et al. \cite{porras00}). The derived luminosity ($6\times 10^3$ L$_{\odot}$) and mass ($20$ M$_{\odot}$) are
typical of the mm cores in $\rho$-Oph. In Fig.~7-middle, G\,173.49-MM4 appears slightly isolated from the group of clearly identified massive 
star-forming clumps
for which $F_{14}>F_8$ and $F_{21}/F_8>2$ (e.g. Lumsden et al. \cite{lumsden02}). G\,173.49+2.42 is therefore a good example of a star-forming
complex in which various mass types of YSO clusters co-exist.

High resolution and high sensitivity observations in mid-IR, mm and cm are needed to differentiate between these two scenarii
by searching for multiple mm core system within each mm clump as seen in $\rho$-Oph and weak free-free emission around high-mass protostars.

\begin{table*}
\caption[]{Summary of the mm clump properties. $A_v$ is derived from $M_{gas}$ in the continuum case
of table 5. Note: $^*$A clump angular size of $24''$ is used.}
\begin{center}
\footnotesize
\begin{tabular}{llllll}
\hline
\hline
Clump   & D     & M                        & $d_{cold}$  & Density                               \\
name    & (kpc) & ($10^2{\rm M}_{\odot}$)  & (pc)        & (${\rm M}_{\odot}~{\rm pc^{-3}}$)  & (mag)   \\
\hline
G\,31.28-MM1  & 5.6 & 12.6 & 0.3 & $89.1\times10^3$        & 186   \\
G\,31.28-MM2  & 5.6 & 3.4  & 0.5 & $4.7\times10^3$         & 50    \\
G\,31.28-MM3  & 5.6 & 1.8  & 0.6$^*$ & $1.6\times10^3$$^*$ & 26    \\
G\,59.78-MM1  & 2.1 & 0.8  & 0.2$^*$ & $17.9\times10^3$$^*$ & 78   \\
G\,59.78-MM2  & 2.1 & 1.0  & 0.1 & $86.9\times10^3$        & 104   \\
G\,173.49-MM1 & 1.8 & 1.2  & 0.1 & $67.9\times10^3$        & 164   \\
G\,173.49-MM2 & 1.8 & 0.3  & 0.1 & $49.4\times10^3$        & 40    \\
G\,173.49-MM3 & 1.8 & 0.3  & 0.2$^*$ & $5.8\times10^3$$^*$ & 35    \\
G\,173.49-MM4 & 1.8 & 0.2  & 0.1 & $139.1\times10^3$       & 28    \\
G\,188.95-MM1 & 2.2 & 0.5  & 0.1 & $95.5\times10^3$        & 48    \\
G\,188.95-MM2 & 2.2 & 1.2  & 0.1 & $117.3\times10^3$       & 114   \\
G\,192.60-MM1 & 2.6 & 2.2  & 0.3 & $17.2\times10^3$        & 150   \\
G\,192.60-MM2 & 2.6 & 3.2  & 0.2 & $89.1\times10^3$        & 237   \\
G\,192.60-MM3 & 2.6 & 1.0  & 0.3$^*$ & $7.3\times10^3$$^*$ & 70    \\
\hline
\end{tabular}
\end{center}
\end{table*}

\section{Conclusions}

Five methanol maser sites have been studied from mid-IR to mm wavelengths.
Each radio-quiet maser site is always associated with a massive ($>50$~M$_{\odot}$),
very luminous ($>10^4$ L$_{\odot}$), deeply embedded ($>40$ mag) molecular clump.
They are all underluminous stellar objects ($L_{total}{\propto}M_{gas}^{0.75}$).
The properties of the maser sites derived from observations and modelling suggest that
they are embedded HMCs in which high-mass stars are in the process of forming.
All these elements demonstrate that the radio-quiet methanol masers trace high-mass star-forming protoclusters in earlier
phases than those detected in the \ion{H}{ii} region surveys.

In addition, colder gas clumps seen only at mm wavelengths are also found near the methanol maser sites. These
might represent an even earlier phase of massive star formation. These results suggest an evolutionary sequence for massive star
formation from a cold clump only seen at mm wavelengths, evolving to a HMC with a two-component SED with peaks at far-IR and mid-IR wavelengths,
and ending with (ultra-compact) \ion{H}{ii} regions very bright at far-IR wavelengths. Alternatively, the cold clumps might be clusters
of less massive ($<8$~M$_{\odot}$) YSOs, in formation near the high-mass star-forming protoclusters.

Finally, the values of the dust grain emissivity index ($\beta$) range between 1.6 and 1.9, which agree with modelling work (Ossenkopf \& Henning
\cite{ossenkopf94}).

\section*{Acknowledgments}

This research was funded by an Australia Research Council Discovery 
Grant. It has also made use of Centre de Donn\'ees
astronomiques de Strasbourg ({\it CDS}), the Digitized Sky Survey produced at the 
Space Telescope Science Institute, the NASA/IPAC Infrared Science Archive operated 
by the Jet Propulsion Laboratory, California Institute of Technology. 
We thank the SEST staff for their help during the
SIMBA observation and data reduction, especially Markus Nielbock
for his data analysis support. The SEST and the Onsala-20m telescope are 
operated by the Swedish National Facility for Radio Astronomy, Onsala Space Observatory 
at Chalmers University of Technology. The Mopra millimetre
radio-telescope is part of the Australia Telescope which is funded
by the Commonwealth of Australia for operation as a National
Facility managed by CSIRO.

\end{document}